\documentclass[usenatbib]{mnras}
\usepackage{graphics,graphicx}
\usepackage{amsmath}
\usepackage{amssymb,latexsym,mathrsfs, bm}
\usepackage{color}
\usepackage{float}
\usepackage{natbib}
\usepackage{times}
\usepackage{pdflscape}
\usepackage{longtable}

\def\eq#1{{Eq.~(\ref{#1})}}

\title[CII intensity mapping]{Constraining the evolution of [CII] intensity 
through  the end stages of reionization}

\author[Padmanabhan]{Hamsa Padmanabhan$^{1,2}$\thanks{Email: 
hamsa@cita.utoronto.ca}
\\ $^{1}$Canadian Institute for Theoretical Astrophysics, 60 St. George St, Toronto, Ontario M5S 3H8, Canada
\\ $^{2}$Institute for Particle Physics and Astrophysics, ETH Zurich, 
Wolfgang-Pauli-Strasse 27, CH-8093 Z\"{u}rich, Switzerland
  }

\date{Accepted ---. Received ---; in original form ---}

\pubyear{2018}

\begin{document}
\label{firstpage}
\pagerange{\pageref{firstpage}--\pageref{lastpage}}
\maketitle

\begin{abstract}
We combine available constraints on the local [CII] 158 $\mu$m line luminosity 
function from galaxy observations \citep{hemmati2017}, with the evolution of 
the star-formation rate density and the recent [CII] intensity mapping 
measurement in \citet[][assuming detection]{pullen2018}, to derive the evolution 
of the [CII] luminosity - halo mass relation over $z \sim 0-6$.  We develop 
convenient fitting forms for the evolution of the [CII] luminosity - halo mass 
relation, and forecast constraints on the [CII] intensity mapping power spectrum 
and its associated uncertainty across redshifts. We predict the sensitivities to 
detect the power spectrum for upcoming PIXIE-,  STARFIRE-, EXCLAIM-, CONCERTO-, 
TIME- and CCAT-p-like surveys, as well as possible future intensity mapping  
observations with the ALMA facility. 
\end{abstract}


\section{Introduction}
Intensity mapping of atomic and molecular lines has emerged as a novel approach 
towards the study of large scale structure. Having been widely used for 
studying neutral hydrogen \citep[e.g.,][]{chang10, masui13, switzer13, 
anderson2018}, it has now emerged as an exciting potential probe of various 
species of ionized and molecular gas in the post-reionization universe  and 
recently, also the epoch of reionization \citep[e.g.,][]{gong2012, lidz2016, 
mashian2015, dumitru2018}. Since intensity mapping is sensitive to the total 
emission from all galaxies, it does not suffer from the magnitude cuts which 
are imposed on galaxy luminosity functions, and may thus offer a powerful tracer 
of the faintest galaxies thought to be responsible for  reionization. Allowing 
for the low luminosity end of the galaxy population to be mapped effectively 
also makes this a complementary technique to galaxy surveys. In addition, 
cross-correlating various intensity mapping datasets provides a possible way of 
probing the large scale structure \citep[e.g.,][]{gong2012}. 

The fine structure line of ionized carbon ([CII]), with a rest wavelength $157.7 
\ \mu$m is thought to be an important tracer of the interstellar medium at the 
late stages of reionization  \citep[e.g.,][]{gong2012, serra2016, li2015, 
crites2017} due to its close association with star forming galaxies. This line 
arises due to a hyperfine transition between the ${}^2 P_{3/2}$ and ${}^2 
P_{1/2}$ states of singly ionized carbon. CII is the dominant coolant in the 
interstellar medium, and with an ionization potential of 11.6 eV, the emission 
line is associated both with neutral and ionized hydrogen regions. The 
continuum foregrounds associated with [CII] emission are also smaller than for 
other emission lines. Above redshifts of about 0.6, the [CII] line becomes the 
brightest in the far-infrared and sub-millimetre part of galaxy spectra. It is 
known that the [CII] emission is closely connected to the star formation rate by 
a power law relation \citep{delooze2014} and thus offers constraints on the 
sources responsible for reionization.  [CII] intensity mapping surveys also have 
the potential to constrain cosmology and models of inflation through 
measurements of primordial non-Gaussianities \citep[e.g.,][]{fonscea2018, 
dizgah2018}.

There are several planned experiments that aim to place constraints on the 
integrated intensity of [CII] over the epoch of reionization, including (i) the 
Cerro Chajnantor Atacama Telescope \citep[CCAT-prime;][]{parshley2018} 
\footnote{http://www.ccatobservatory.org/docs/ccat-technical-memos/ccatp\_im\_v2.pdf} 
which plans to trace [CII] over $z \sim 5-8$, probing the late stages of 
reionization, (ii) the Tomographic Intensity Mapping Experiment \citep[TIME; 
e.g.,][]{crites2014, crites2017} and (iii) the CarbON CII line in 
post-rEionization and ReionizaTiOn \citep[CONCERTO;][]{serra2016, lagache2018}  
experiment\footnote{https://people.lam.fr/lagache.guilaine/CONCERTO.html} which 
plans to observe the evolution of CII over $z \sim 4.5 - 8.5$.  Above $z \sim 
4$, the [CII] line redshifts into the millimetre wavelengths, where the Atacama 
Large Millimetre Array (ALMA) facility can make observations in the single dish 
mode; the ALMA  bands 7 and 8 enable [CII] emission observations from $z \sim 2.8 
- 5$.

Much of the present data available in the context of [CII] observations  at the 
epoch of reionization relies on individual galaxies, imaged e.g. with the ALMA 
telescope \citep[][]{capak2015, smit2018, pentericci2016}. On the other hand, 
local galaxies have been used to compute the [CII] luminosity function at $z \sim 
0$ \citep{hemmati2017}. At redshifts $z \sim 2$, observations of [CII] offer 
interesting prospects towards constraining the peak of the star-formation rate 
density of the  universe, however, these wavelengths are not accessible from 
the ground. Nevertheless, there are interesting prospects for balloon-based 
observations of [CII] at intermediate redshifts, $z \sim 1 - 1.5$ 
\citep{uzgil2014} as well as from upcoming studies of CMB spectral distortions 
\citep[e.g.,][]{kogut2011, hernandez2017}. Recently, the first tentative 
detection of the  integrated [CII]  intensity at $z \sim 2.6$ was reported in 
\citet{pullen2018} using Planck intensity maps cross-correlated with BOSS 
quasars and CMASS galaxies from SDSS-III.  The main observable in a [CII] 
intensity mapping survey is the power spectrum of intensity fluctuations, 
$P_{\rm CII} (k,z)$ as a function of scale and redshift. The astrophysical 
component of the power spectrum is encoded by the relation between the [CII] 
luminosity (which is used in computing the integrated emission intensity), and 
the underlying halo mass. Several theoretical and simulation approaches have 
been used to place constraints on this relation and the power spectrum of [CII] 
fluctuations up to  the mid- to the end stages of reionization 
\citep[e.g.,][]{gong2012, yue2015, serra2016, silva2015, dumitru2018}.

In this paper, we adopt a data-driven, halo model based approach to 
constraining the evolution of the [CII] luminosity - halo mass relation across 
redshifts with the help of the available data. We use the technique of 
abundance matching to derive constraints on the local [CII] luminosity - halo mass 
relation from the observations of the $z \sim 0$ [CII] luminosity function 
\citep{hemmati2017}. We combine this information with the recent high-redshift 
constraints on the [CII] intensity \citep[][assuming the CII 
detection]{pullen2018} at $z \sim 2.6$ to model the evolution of this relation 
at higher redshifts, and find the predictions to be to consistent with the 
currently available observational limits from galaxy data at $z \sim 4-6$ 
\citep{matsuda2015, swinbank2012, aravena2016}. We use the [CII] luminosity - 
halo mass relation thus derived to calculate the power spectrum of intensity 
fluctuations across $z \sim 0-6$, and forecast its measurement sensitivity by 
future intensity mapping experiments.

The plan of the paper is as follows: In Sec. \ref{sec:model}, we describe the 
current [CII] data  used for constraining the model parameters and their 
evolution. We also describe the abundance matching procedure and the parameter 
constraints with their associated uncertainties. In the following section (Sec. 
\ref{sec:powspec}), we use the resultant [CII] luminosity - halo mass relation to 
predict the power spectra of intensity fluctuations, both at low and high 
redshifts (up to $z \sim 6$). In Sec. \ref{sec:forecasts}, we use the results 
of the model with the parameters of several planned or upcoming [CII] intensity 
mapping experiments to place sensitivity forecasts on the expected [CII] 
auto-correlation signal. We summarize our results and discuss future prospects 
in Sec. \ref{sec:conclusions}. Throughout this work, we adopt a $\Lambda$CDM 
cosmology with the following parameters: $h = 0.71, \Omega_b = 0.046, \Omega_m 
= 0.281, \sigma_8 = 0.8, \Omega_{\Lambda} = 0.719, n_s = 0.963$.

\begin{figure}
\includegraphics[scale = 0.6, width = \columnwidth]{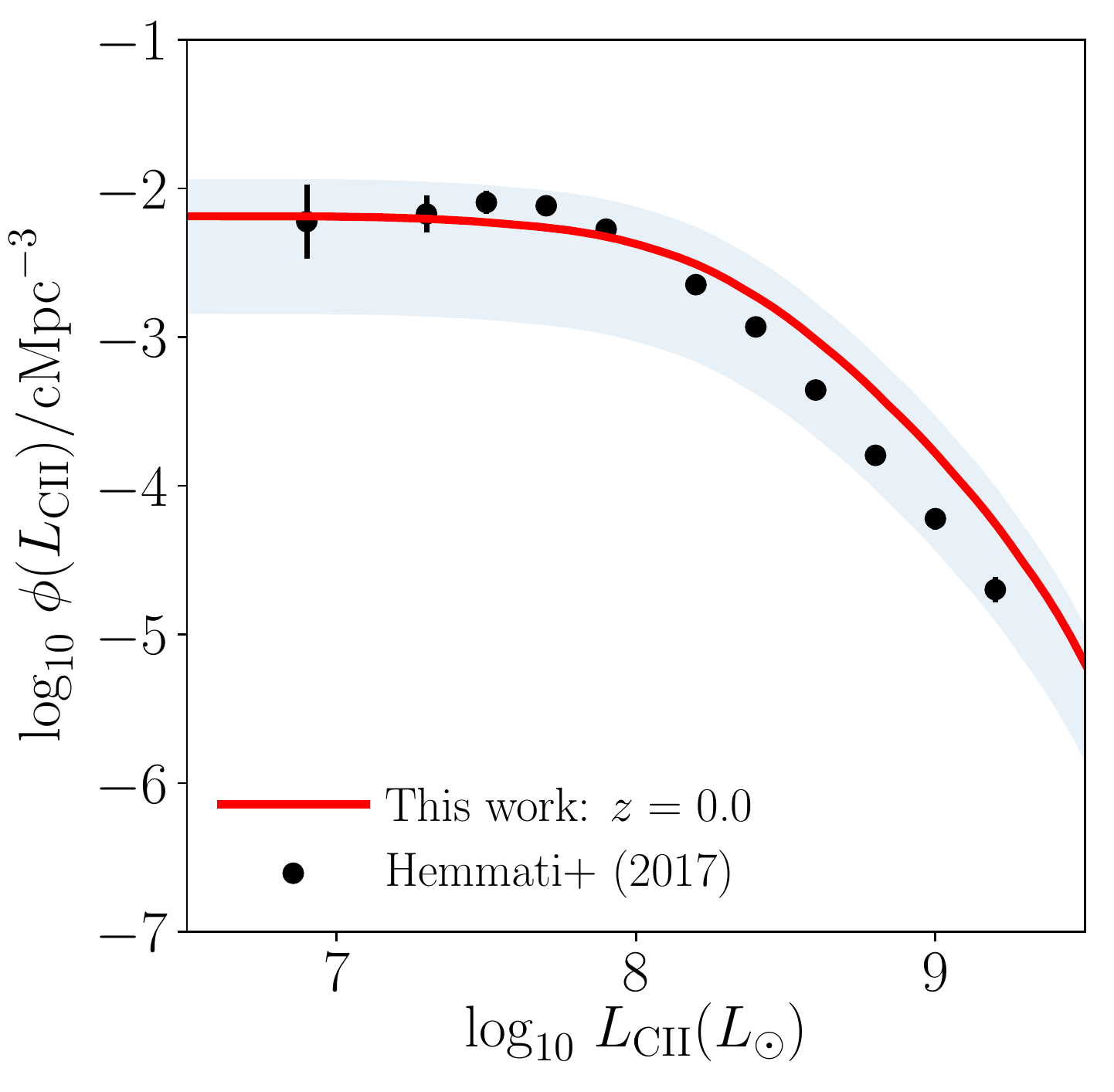} 
\caption{[CII] luminosity function at $z \sim 0$, along with the observational 
constraints \citep{hemmati2017} used in empirically deriving the $L_{\rm CII} - 
M$ relation.}
\label{fig:ciilowz}
\end{figure}

\section{Modelling the occupation of CII}
\label{sec:model}

The main observable in [CII] intensity mapping is the power spectrum of 
fluctuations, denoted by $P_{\rm CII} (k,z)$. This quantity is related to the 
intensity of the observed emission, given by:
\begin{equation}
 I_{\nu, {\rm CII}} = \frac{c}{4 \pi} \int_0^{\infty} dz' 
\frac{\epsilon[\nu_{\rm obs} (1 + z')]}{H(z') (1 + z')^4}
\end{equation} 
where the emissivity, $\epsilon(\nu)$ is given by:
\begin{equation}
 \epsilon(\nu, z) = \delta_D(\nu - \nu_{\rm CII}) (1 + z)^3 \int_{M_{\rm min, 
CII}}^{\infty} dM \frac{dn}{dM} L_{\rm CII}(M,z)
\end{equation} 
where $L_{\rm CII}$ represents the luminosity of CII-emitting galaxies as a 
function of their halo mass $M$ and redshift $z$, and $M_{\rm min, CII}$ stands 
for the minimum halo mass associated with CII-emitting galaxies.
Thus the intensity of emission becomes:
\begin{equation}
I_{\nu, {\rm CII}} = \frac{c}{4 \pi} \frac{1}{\nu_{\rm CII} H(z_{\rm CII})}  
\int_{M_{\rm min, CII}}^{\infty} dM \frac{dn}{dM} L_{\rm CII}(M,z)
\label{CIIspint}
\end{equation} 
The above expression thus depends on the relation between the [CII] luminosity 
and halo mass at different redshifts. Various approaches in the literature have 
been used to constrain this relation from hydrodynamical simulations and 
semi-analytical techniques. Here, we develop a data-driven, halo model 
framework for modelling this relation, following similar approaches for 
stellar-halo mass relations \citep[e.g.,][]{behroozi2013,moster2013}, the HI - 
halo mass relation \citep[e.g.,][]{hpar2017, hpgk2017, hparaa2017} and the 
evolution of the CO luminosity - halo mass relation \citep[e.g.,][]{hpco}.

The method outlined here relies on the technique of \textit{abundance 
matching}, in which the relative abundances of [CII]-luminous galaxies and their 
associated host haloes are assumed to follow a one-to-one correspondence. In 
other words, the brightest [CII] emitting galaxies are assumed to populate the 
most massive dark matter haloes. The is equivalent to the assumption that:
\begin{equation}
  \int_{M (L_{\rm CII})}^{\infty} \frac{dn}{ d \log_{10} M'} \ d \log_{10} M' = 
\int_{L_{\rm CII}}^{\infty} \phi(L_{\rm CII}) \ d \log_{10} L_{\rm CII}
  \label{eqn:abmatchcii}
\end{equation}
where $dn / d \log_{10} M$ is the number density of dark matter haloes with 
logarithmic
masses between $\log_{10} M$ and $\log_{10} (M$ + $dM)$, and $\phi(L_{\rm 
CII})$ is the
corresponding number density of [CII]-luminous galaxies in logarithmic luminosity 
bins.

At redshift zero, we use the recent available constraints on the local [CII] 
luminosity function observed with the Herschel PACS observations of the 
Luminous Infrared Galaxies in the Great Observatories All-sky LIRG Survey 
\citep{hemmati2017} to constrain the local [CII] luminosity - halo mass relation. 
Motivated by approaches to model the atomic hydrogen gas, in e.g., 
\citet{hparaa2017}, we use an $L_{\rm CII} - M$ relation at $z \sim 0$ having a 
power law form with an exponential cutoff:
\begin{equation}
L_{\rm CII} (M, z = 0) = \left(\frac{M}{M_1}\right)^{\beta} \exp(-N_1/M)
\end{equation}
with the three free parameters, $M_1$, $\beta$ and $N_1$ for the two 
characteristic mass scales and the slope of the relation.

\begin{figure}
\includegraphics[scale = 0.6, width = \columnwidth]{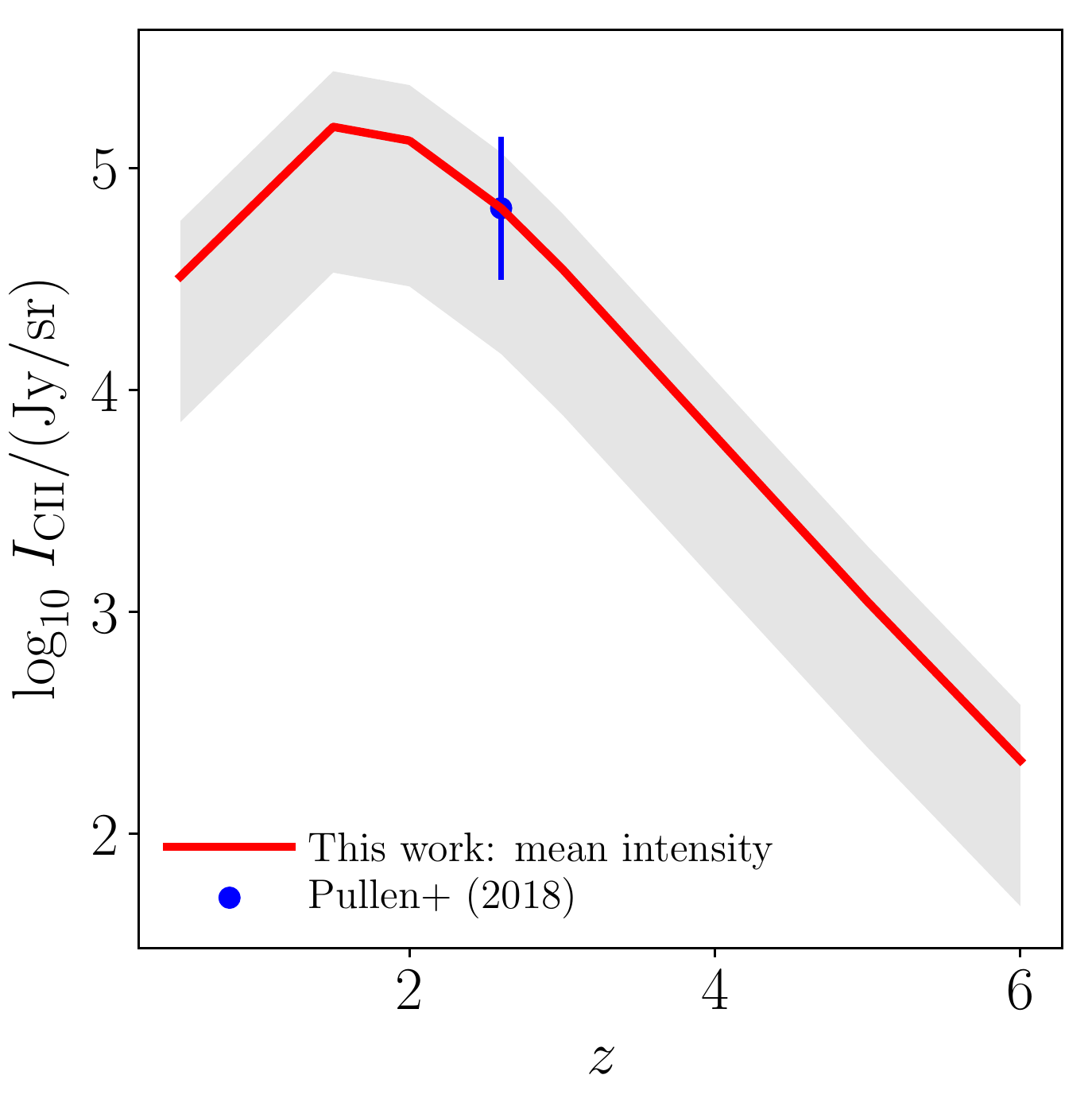} 
\caption{Evolution of mean intensity $I_{\rm CII}$ across redshifts from the 
best-fitting model parameters. Also shown is the intensity mapping constraint 
at $z \sim 2.6$ from the results of \citet{pullen2018}.}
\label{fig:iciiallz}
\end{figure}

We use the observed [CII] luminosity function [the raw data points in 
\citet{hemmati2017}] abundance-matched to the Sheth-Tormen \citep{sheth2002} 
halo mass function in order to derive constraints on the parameters of the 
$L_{\rm CII} - M$ relation from
Equation~(\ref{eqn:abmatchcii}). This  gives the following best-fitting and 
error values for the free parameters:
\begin{eqnarray}
M_1 &=& (2.39 \pm 1.86) \times 10^{-5}; \  N_1 = (4.19 \pm 3.27) \times 
10^{11}; \nonumber \\
 & & \beta = 0.49 \pm 0.38
\end{eqnarray}

Fig. \ref{fig:ciilowz} shows the luminosity function at $z \sim 0$ obtained 
from the best-fitting $L_{\rm CII} - M$ relation thus obtained, and its 
associated uncertainty. Also shown are the data points from \citet{hemmati2017} 
used in deriving this relation. 

At high redshifts, the following datasets are available from candidate [CII] 
galaxies and blind searches:

\begin{enumerate}
\item At $z \sim 4.4$, \citet{swinbank2012} provide lower limits on the 
cumulative [CII]  luminosity function based on observations of two ALMA candidate 
sub-millimetre galaxies (SMGs). 

\item \citet{matsuda2015} obtain upper limits on the cumulative [CII] luminosity 
function at $z \sim 4.5$, based on a blind search using ALMA archival data.

\item \citet{aravena2016} obtain upper limits on the cumulative [CII] luminosity 
function at from the ASPECS blind survey $z \sim 6-8$ from  candidate  
[CII]-emitters.

\end{enumerate}

The first constraints on the integrated  [CII] emission, $I_{\nu, {\rm CII}}$  at 
$z \sim 2.6$ have been placed by \citet{pullen2018} using cross-power spectra 
between high-frequency Planck intensity maps, spectroscopic quasars from 
BOSS-DR12 and CMASS galaxies from SDSS-III, finding a tentative [CII] intensity 
measurement of $I_{\nu, {\rm CII}} = 6.6^{+5.0}_{-4.8} \times$ 10$^4$ Jy/sr at 
95\% confidence.

To propagate the empirically derived $L_{\rm CII} - M$  relation to higher 
redshifts, we use the observed  evolution of the star formation rate density 
\citep[SFRD;][]{madau2014}. This is consistent with the results of 
\citet{hemmati2017}, who find the observed [CII] luminosity function to evolve 
following that of the SFR \citep{behroozi2013}, and \citet{lagache2018a} whose 
simulations do not find evidence for significant evolution in the $L_{\rm CII}$ 
- SFR relation. The results of \citet{pentericci2016} and \citet{aravena2016} 
also find that the observations in the late stages of reionization closely 
follow the SFR-[CII] scaling, expressed in a power law form as $L_{\rm CII} = 
{\rm SFR}^{\alpha}$. Hence, the high-$z$ [CII]-halo mass relation can be 
expressed as :
\begin{equation}
L_{\rm CII}(M,z) = \left(\frac{M}{M_1}\right)^{\beta} \exp(-N_1/M) 
\left(\frac{(1+z)^{2.7}}{1 + [(1+z)/2.9)]^{5.6}} \right)^{\alpha}
\label{lciihighz}
\end{equation}

\begin{figure}
\includegraphics[scale = 0.6, width = \columnwidth]{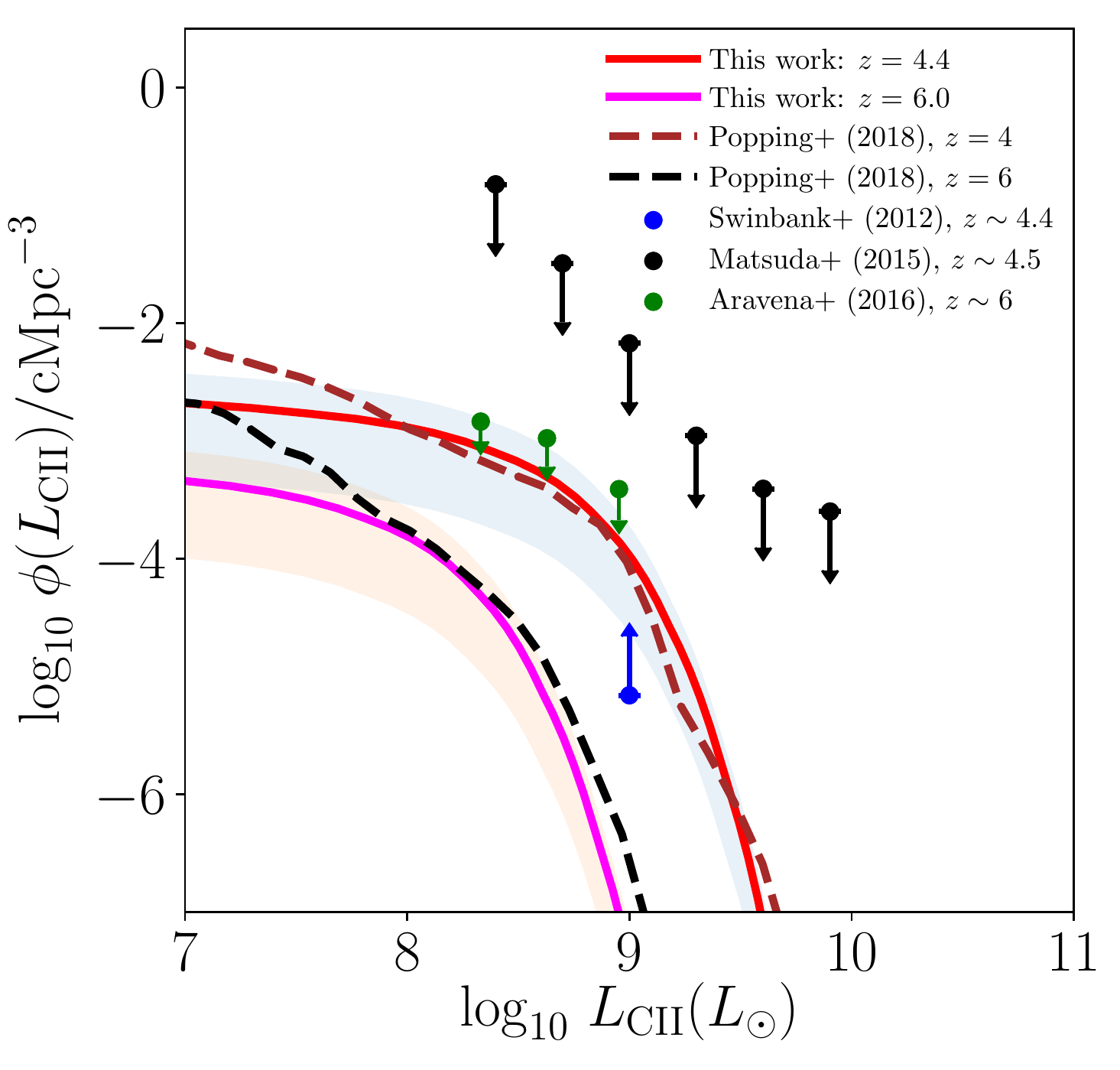}  
\caption{[CII] luminosity function from the best-fitting model at higher 
redshifts ($z \sim 4.4$ and $z \sim 6$), along with the currently available 
limits at $z \sim 4.4$ \citep{swinbank2012, matsuda2015} and $ z\sim 6$ 
\citep{aravena2016}. Also shown are the results from the semi-analytic 
modelling of \citet{popping2018} at $z \sim 4$ and $z \sim 6$.}
\label{fig:ciilumfunc46}
\end{figure}

The value of $\alpha$ is constrained by fitting the intensity mapping 
measurement of \citet{pullen2018} at $z \sim 2.6$ using \eq{CIIspint}. 
\footnote{This measurement is also sensitive to the minimum halo mass $M_{\rm 
min, CII}$ used in the calculation of $I_{\nu, \rm CII}$. Throughout this work, 
we use $M_{\rm min, CII} = 10^{10} h^{-1} M_{\odot}$ for consistency, which 
also gives a good fit to the intensity mapping data for sensible values of the 
evolution parameter $\alpha$.} 
This leads to $\alpha = 1.79 \pm 0.30$; the intensity mapping data may favor 
values consistent with (though slightly higher than) those predicted by the 
simulations of \citet{lagache2018a}.The present data do not constrain a 
mass-dependent redshift evolution of the $L_{\rm CII} - M$ relation, but only 
the component arising from the evolution of the star formation rate 
density.\footnote{Note that this makes the implicit assumption that the SFRD is 
separable, i.e. SFR $(M,z) = f(M)g(z)$. Although this separability may not be 
strictly true in general, this functional form is found to provide a good fit 
to the available [CII] data (and is also supported by the fact that the [CII] 
luminosity functions thus derived closely resemble those from semi-analytic 
models invoking more detailed sub-grid treatments of SFR and ISM physics across 
redshifts, e.g., in Fig. \ref{fig:ciilumfunc46}.} The data also do not 
constrain the redshift evolution of the parameter $\alpha$ itself, though 
several studies report the values for different galaxy samples across redshifts 
\citep[e.g.,][]{delooze2011}. Such a redshift dependence  may also be suggested 
by the results of simulations \citep{lagache2018a}. 

{Keeping in mind that the present study does not consider instrumental systematic effects or foreground contamination, the quoted errors are conservative estimates: the fitting errors are added in quadrature to the observational uncertainties from \citet{hemmati2017} for the case of $z \sim 0$. For the high-$z$ parameters, the upper and lower limits of the \citet{pullen2018} measurement are used along with the fitting errors to estimate the parameter uncertainties.}

The predicted evolution of the mean intensity of [CII] emission, $I_{\nu, \rm 
CII}$ across redshifts is shown in Fig. \ref{fig:iciiallz} along with the data 
point at $z \sim 2.6$ from \citet{pullen2018}.
The resultant $L_{\rm CII} - M$ relation is also found to be consistent with 
the limits from the observations \citep{matsuda2015, swinbank2012, aravena2016} 
at higher redshifts $z \sim 4-6$, as shown in Fig. \ref{fig:ciilumfunc46}. Also 
shown are the results form the semi-analytical modelling of \citet{popping2018} 
at $z \sim 4$ and $z \sim 6$, which uses a sub-grid treatment of SFR and ISM 
physics across redshifts.

\section{Resultant [CII]-halo mass relation and power spectrum}
\label{sec:powspec}

\begin{figure}
\includegraphics[scale = 0.6, width = \columnwidth]{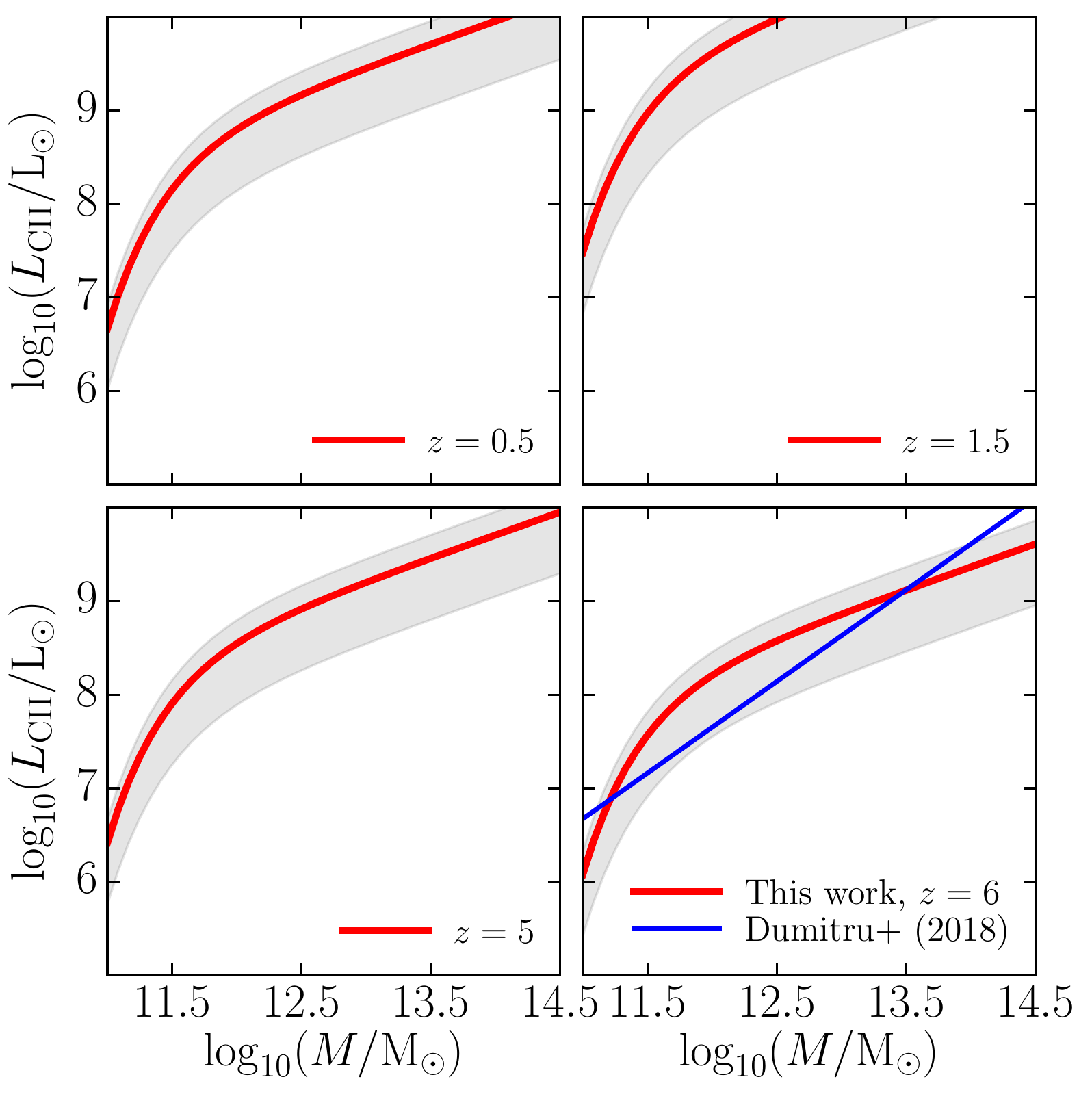} 
\caption{The $L_{\rm CII} - M$ relation for $z = 0.5, 1.5, 5$ and 6 from the 
best-fitting model. The associated uncertainties are shown by the grey bands. The $L_{\rm CII} - M$ relation from the simulations of 
\citet{dumitru2018} is overplotted at $z \sim 6$ for comparison.}
\label{fig:allz}
\end{figure}

\begin{figure}
\includegraphics[scale = 0.6, width = \columnwidth]{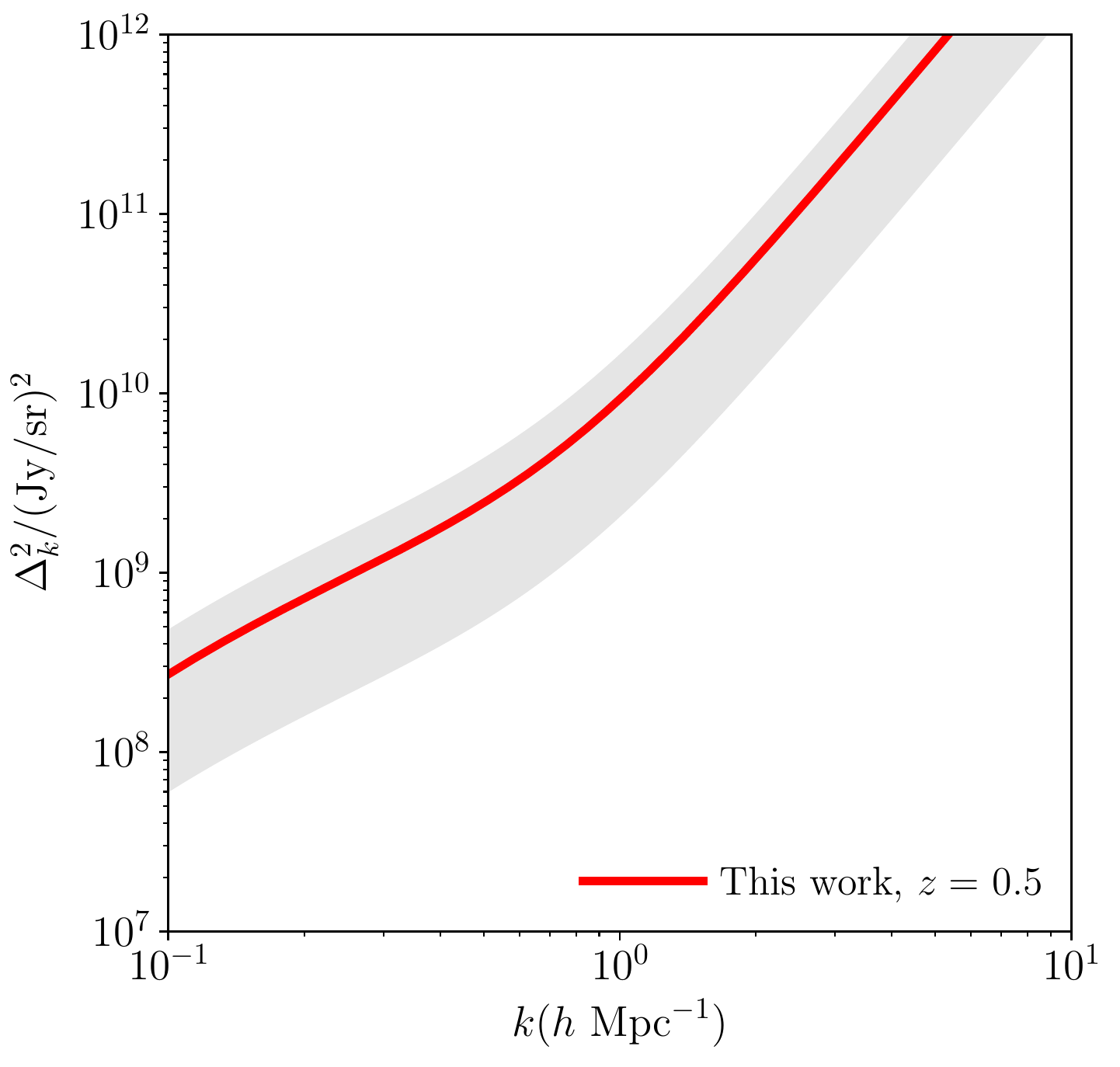} 
\includegraphics[scale = 0.6, width = \columnwidth]{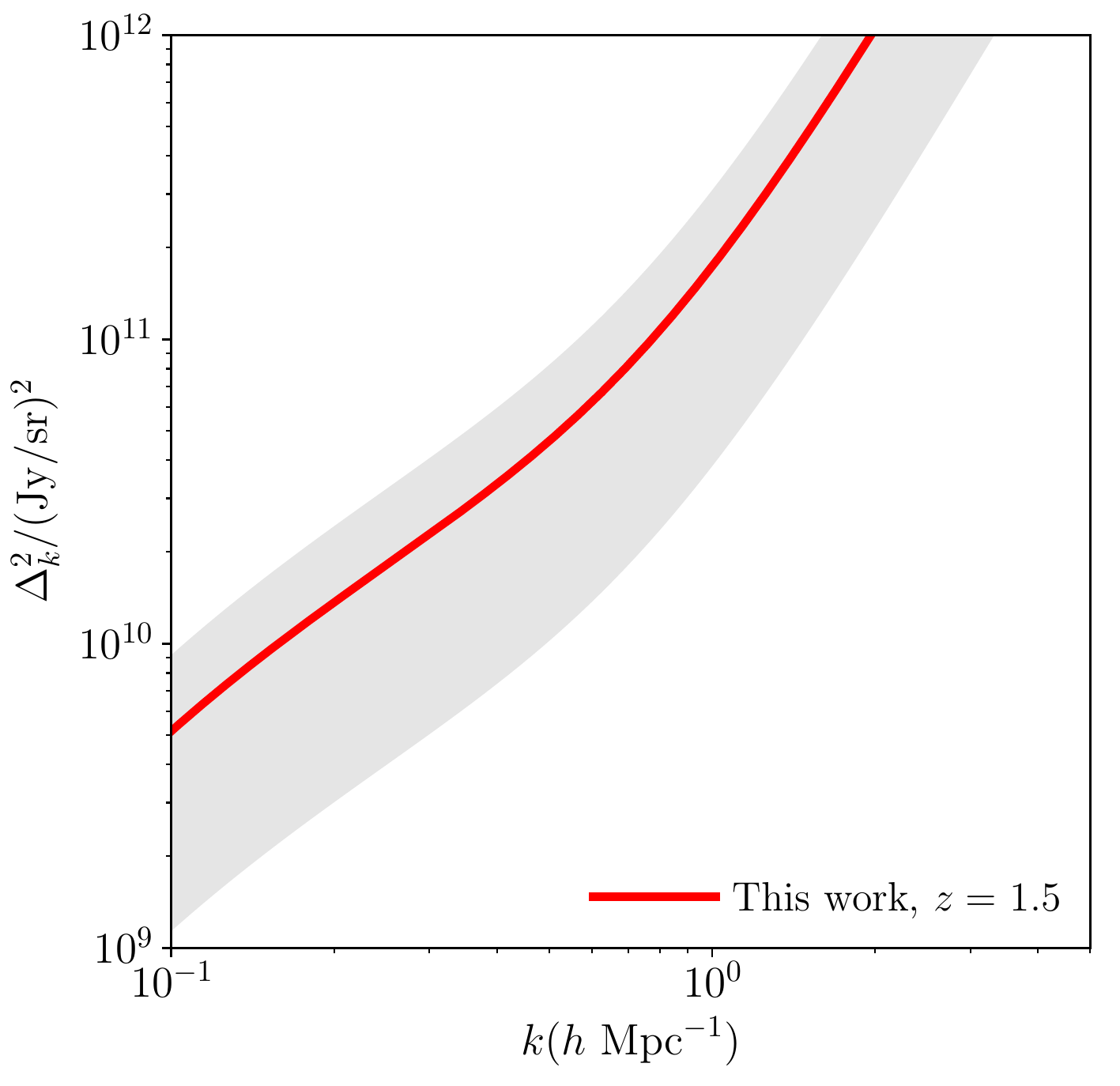} 
\caption{[CII] intensity power spectra at $z \sim 0.5, 1.5$ from the empirical 
relation. The associated uncertainties are shown by the grey 
bands.}
\label{fig:powspec01}
\end{figure}

\begin{figure}
\includegraphics[scale = 0.6, width = \columnwidth]{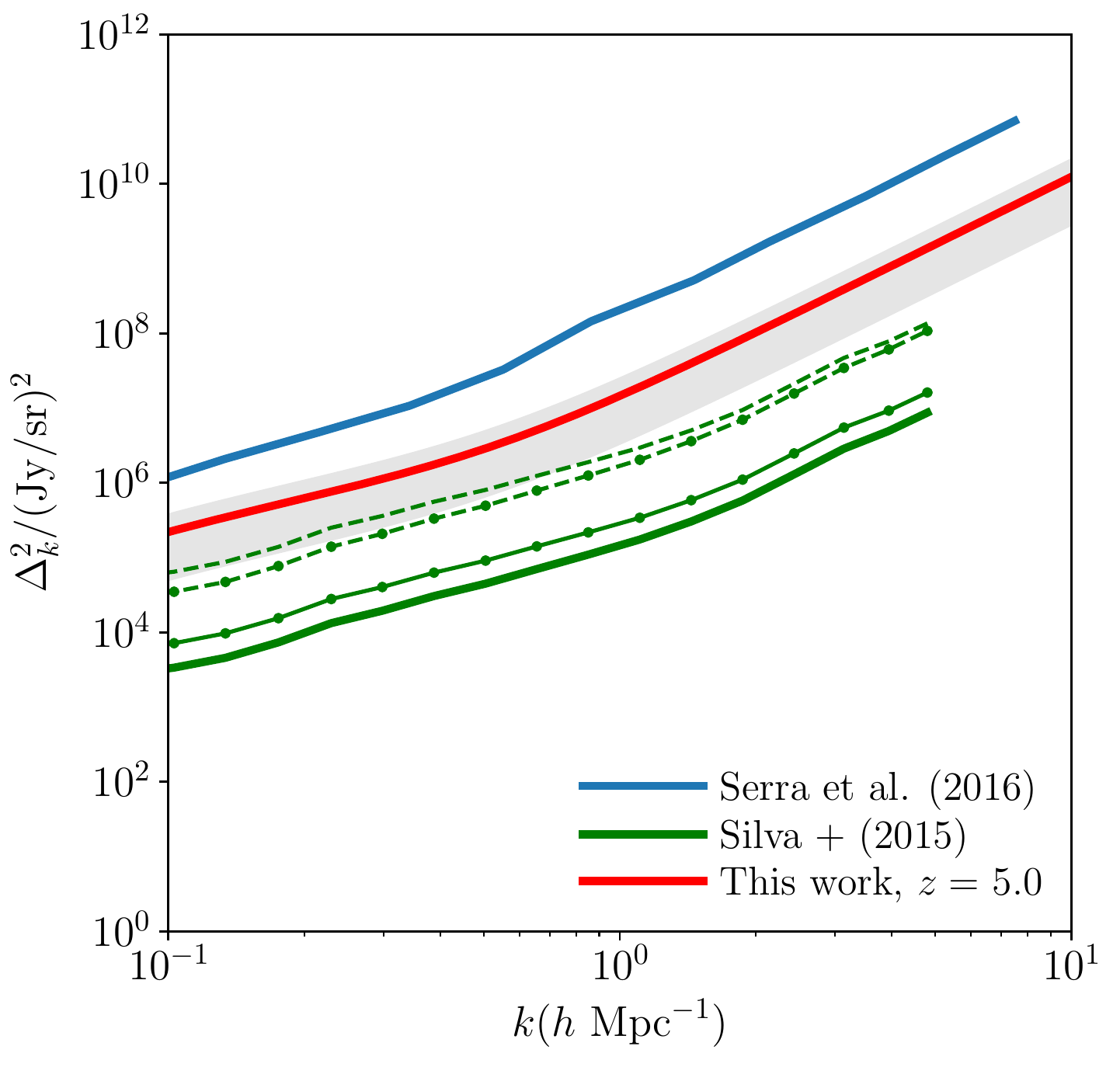} 
\includegraphics[scale = 0.6, width = \columnwidth]{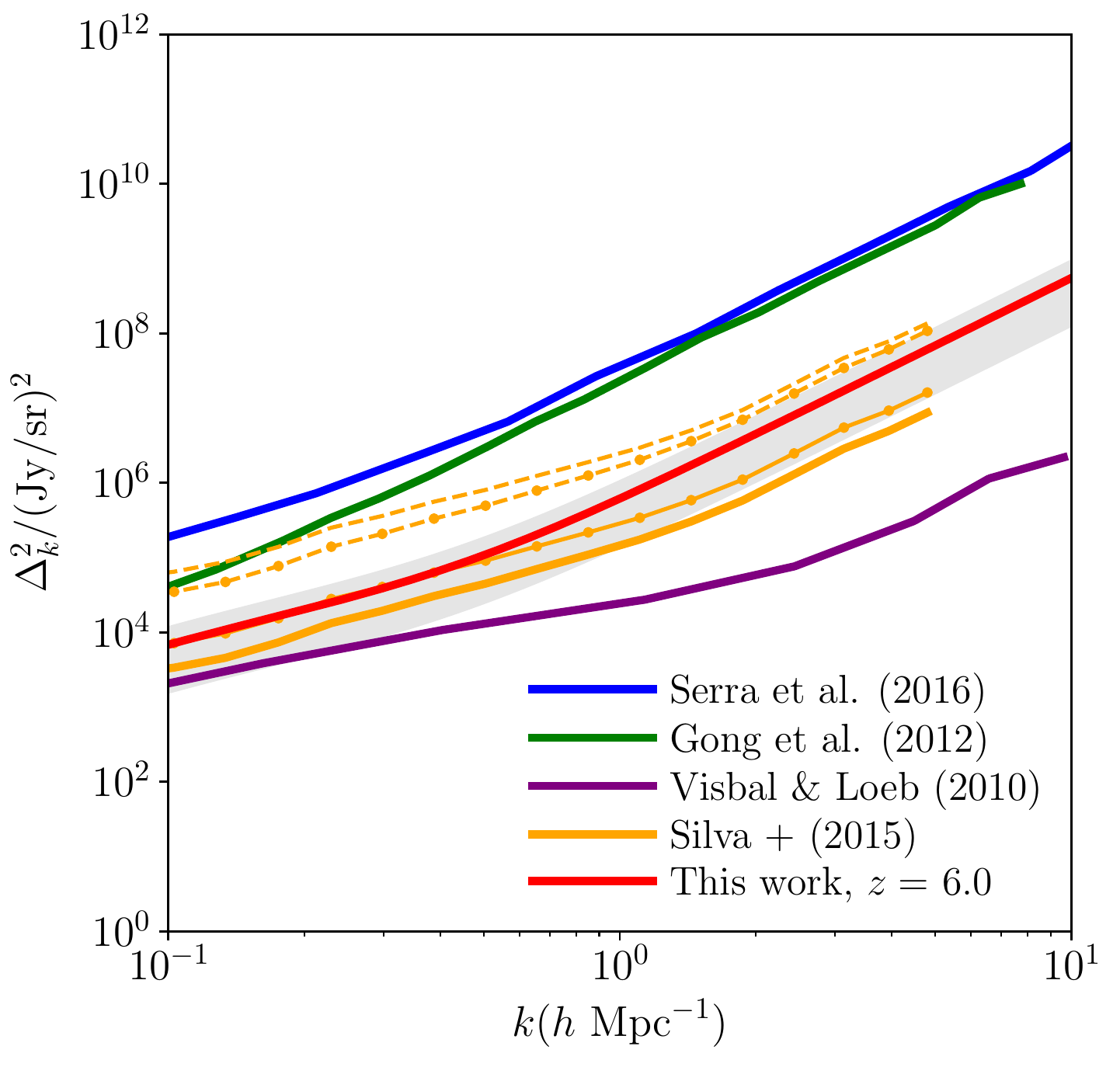} 
\caption{[CII] intensity power spectra at $z \sim 5, 6$ from the present work 
compared to other model and simulation predictions from the literature, from 
\citet{serra2016, gong2012, visbal2010, silva2015}. Note that the 
\citet{silva2015} models are for $ z \sim 6.7$. The different linestyles for 
the \citet{silva2015} denote the four different models m1, m2, m3 and m4 in 
that work.}
\label{fig:powspec56}
\end{figure}

The [CII] luminosity - halo mass relation from the present data can thus be 
described by the functional form in \eq{lciihighz}, with the best-fitting 
parameters:
\begin{eqnarray}
& & M_1 = (2.39 \pm 1.86) \times 10^{-5}; \  N_1 = (4.19 \pm 3.27) \times 
10^{11}; \nonumber \\
 & & \beta = 0.49 \pm 0.38; \  \alpha = 1.79 \pm 0.30
\end{eqnarray}
The relation for redshifts 0.5, 1.5, 5 and 6 is plotted on the panels of Fig. \ref{fig:allz} along with the estimated uncertainties (shown by the grey bands {\footnote{The maximum relative errors in the fitted parameters are used to derive conservative uncertainty estimates for the $L_{\rm CII} - M$ relation and power spectrum. Possible uncertainties in the $M_{\rm min, CII}$ parameter are not folded into the present estimates.}}).
 These can now be used to derive the predicted [CII] power spectrum of intensity 
fluctuations using the relations:
\begin{equation}
 P_{\rm shot}(z) = \frac{\int_{M_{\rm min, CII}}^{\infty} dM (dn/dM) L_{\rm 
CII} (M,z)^2}{\left(\int_{M_{\rm min, CII}}^{\infty} dM (dn/dM) L_{\rm CII} 
(M,z)\right)^2}
\end{equation}
representing the shot noise component, and 
\begin{equation}
 b_{\rm CII}(z) = \frac{\int_{M_{\rm min, CII}}^{\infty} dM (dn/dM) L_{\rm CII} 
(M,z) b(M,z)}{\int_{M_{\rm min, CII}}^{\infty} dM (dn/dM) L_{\rm CII} (M,z)}
\end{equation} 
representing the clustering. In the above expression, $b(M,z)$ denotes the dark 
matter halo bias following \citet{scoccimarro2001}. These can be combined to 
obtain the full power spectrum:
\begin{equation}
 P_{\rm CII}(k,z) =  I_{\nu, \rm CII} (z)^2 [b_{\rm CII}(z)^2 P_{\rm lin}(k,z) 
+ P_{\rm shot}(z)]
\end{equation} 
The power spectra of intensity fluctuations thus derived are plotted for $z = 
\{0.5, 1.5, 5, 6\}$ in the panels of Figs. \ref{fig:powspec01} and 
\ref{fig:powspec56}, in units of (Jy/sr)$^2$).  Also plotted for the higher redshift panels are the semi-empirical and 
simulation estimates for the same redshifts obtained in the literature:
 \begin{enumerate}
 \item \citet{visbal2010} present an analytical formalism to calculate the 
cross-correlation power spectra for different emission lines including OI, OII, 
OIII, NII and CO transitions.
 
 \item \citet{gong2012} obtain the [CII] line intensity fluctuations over $z \sim 
6-8$ by modelling the various radiative processes in the ISM both analytically 
and numerically.
 
 \item \citet{silva2015} use four different models, m1, m2, m3 and m4 
corresponding to different parameterizations of the [CII] luminosity to SFR 
relation coupled to results from simulations. 
 
 \item \citet{serra2016} present a framework based on measurements of the 
cosmic infrared background (CIB) to compute the intensity mapping power spectra 
of multiple far infra-red cooling lines in the ISM, including [CII], [NII], [OI] and 
the CO transitions.
 
 \end{enumerate}
 
Although there is considerable variation in the power spectra values in the 
literature, we note that the predicted values are also sensitive to the choice 
of the parameter $M_{\rm min, CII}$ as discussed in the previous section. 
Nevertheless, these plots provide an estimate of the effect of the intensity 
mapping measurement, if confirmed, on the sensitivities of future [CII] 
measurements at post-reionization epochs. We explore this more fully in the 
following section.

\section{Sensitivity forecasts}
\label{sec:forecasts}

In this section, we use the predicted evolution of the power spectra of [CII] 
intensity fluctuations to place constraints on the SNR expected from current 
and future experiments targeting [CII] over post-reionization epochs. We consider 
configurations resembling the following [CII] experiments in this work, with 
parameters as provided in Table \ref{table:ciisurveys}:

\begin{enumerate}

\item A Primordial Inflation Explorer \citep[PIXIE;][]{kogut2011, 
kogut2014}-like mission which aims to measure spectral distortions of the CMB 
over a broad range of frequencies. This experiment will be able to detect [CII] 
intensity fluctuations at $z \sim 0.05 - 11.7$, and is suitable for wide field 
intensity mapping surveys, including in cross-correlation with galaxy data 
\citep[e.g.,][]{uzgil2014, switzer2017, pullen2018}. This survey can be assumed 
to cover the full sky area (E. Switzer, private communication); we restrict to 
$2 \pi$ sr, corresponding to about 20000 deg$^2$ in order to avoid the Galactic 
plane.

\item An upcoming Spectroscopic Terahertz Airborne Receiver for Far-InfraRed 
Exploration \citep[STARFIRE;][]{aguirre2015, 
hailey2018}\footnote{
https://asd.gsfc.nasa.gov/conferences/FIR/posters/Aguirre\_STARFIRE.pdf}-like 
experiment, a balloon-borne, far-infrared spectroscopic array with  tomographic 
sensitivity to [CII] over $z \sim 0.5 - 1.6$, planned to be hosted on a 2.5 m 
aperture telescope.

\item An EXperiment for Cryogenic Large-Aperture Intensity Mapping 
(EXCLAIM)-like configuration (E. Switzer and A. Pullen, private communication) 
for CO and [CII] at redshifts $z \sim 3$, to be hosted on a 0.74 m aperture 
telescope.

\item The CONCERTO experiment aims to map CII evolution over $z \sim 4.5-8.5$ 
using the  12-m APEX telescope at Chajnantor in Chile \citep{lagache2018a, 
serra2016}. Parameters for a CONCERTO-like configuration are based largely on 
the recent analysis of \citet{dumitru2018}.

\item A Tomographic Intensity Mapping Experiment 
\citep[TIME;][]{crites2017}-like configuration which aims to measure the [CII] 
line intensity emission over the redshift range $5 < z < 9$. We consider a 
survey area of 78' by 0.5' with a 12-m  aperture telescope, and other 
parameters of the TIME-Pilot experiment following the discussions in 
\citet{crites2014}.

\item The Cerro Chajnantor Atacama Telescope (CCAT)-prime experiment 
\citep{parshley2018} with the Prime-Cam instrument \citep{vavagiakis2018} and 
P-Spec imaging spectrometer, plans to probe the late stages of reionization 
using [CII] line emission over $z \sim 5-8$. The parameters adopted for a  
CCAT-p-like configuration are based on the discussions in the document, 
https://www.ccatobservatory.org/docs/pdfs/Draft\_CCAT-p.prospectus.170809.pdf.

\item Finally, we consider a [CII] intensity mapping survey with the ALMA 
telescope \citep[e.g.][]{carilli2018} based on the ALMACAL experiment 
\citep[e.g.,][]{klitsch2018}  targeting [CII] line emission over $z \sim 5-9$. 
Such surveys have been discussed, e.g., in the context of the ALMA 
Spectroscopic Survey in the Hubble Deep Field \citep[ASPECS;][]{walter2014, 
carilli2016}. The parameters used here correspond to a survey using the ALMA 
array over 1000 hours of observation targeting 5 arcmin$^2$.

\end{enumerate}

\begin{figure}
\includegraphics[scale = 0.6, width = \columnwidth]{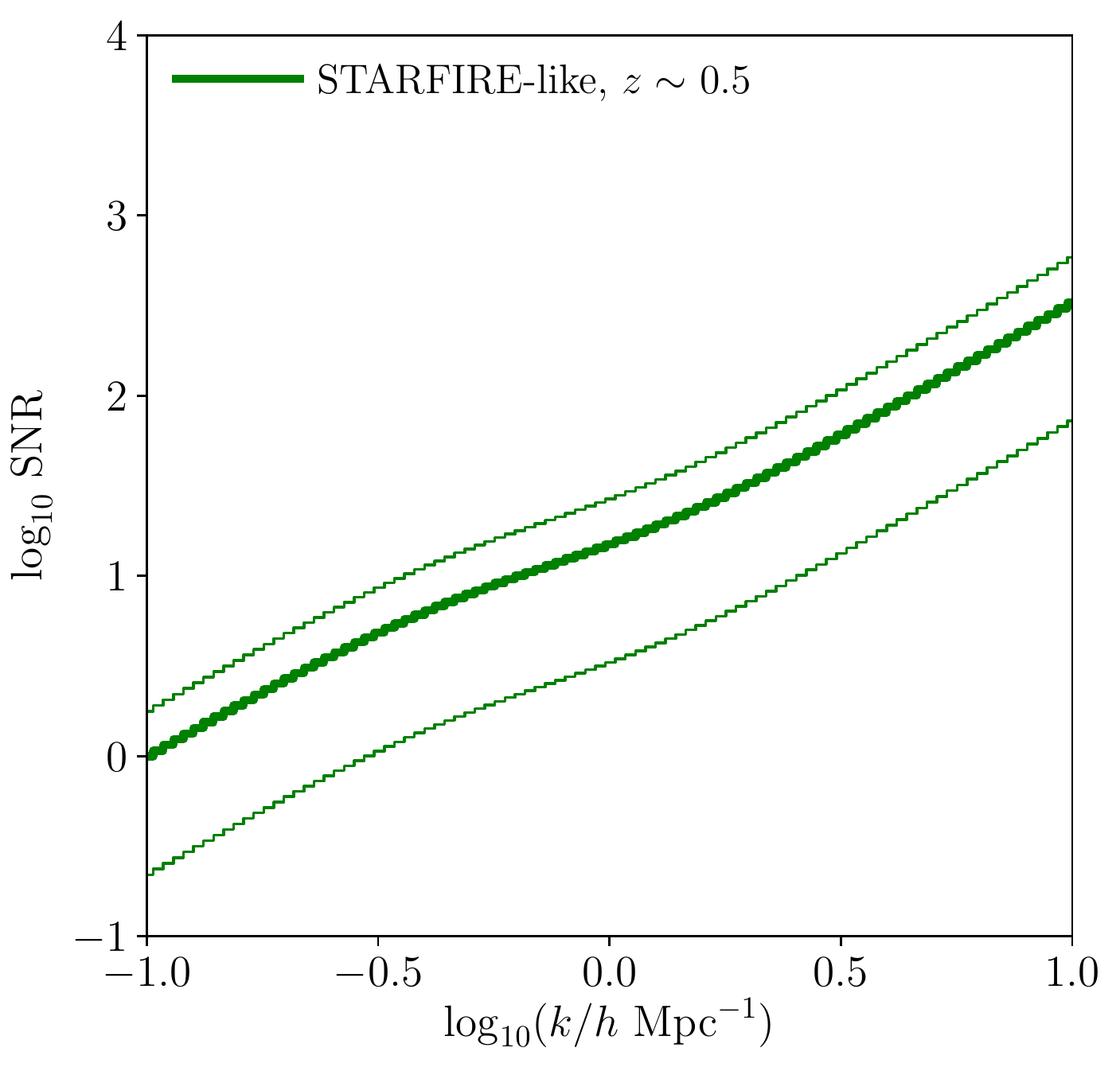} 
\includegraphics[scale = 0.6, width = \columnwidth]{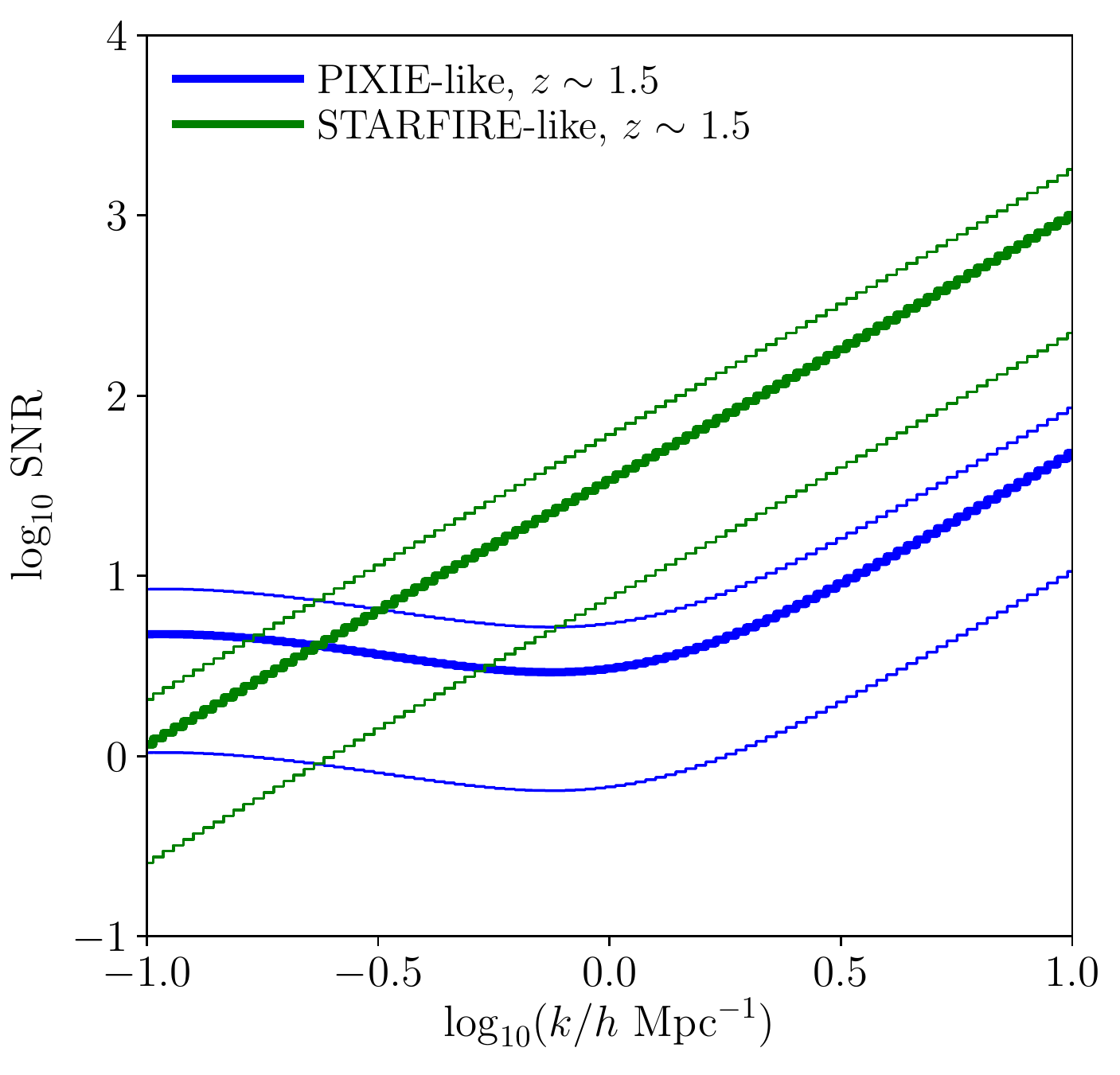} 
\caption{Signal-to-noise ratios at $z \sim 0.5$ (top panel) and $z \sim 1.5$ 
(lower panel) from the empirical relation and the parameters of the PIXIE- and 
STARFIRE-like configurations. The associated uncertainties in the mean value of 
the signal are indicated by the thinner steps.}
\label{fig:snr01}
\end{figure}

\begin{figure}
\includegraphics[scale = 0.6, width = \columnwidth]{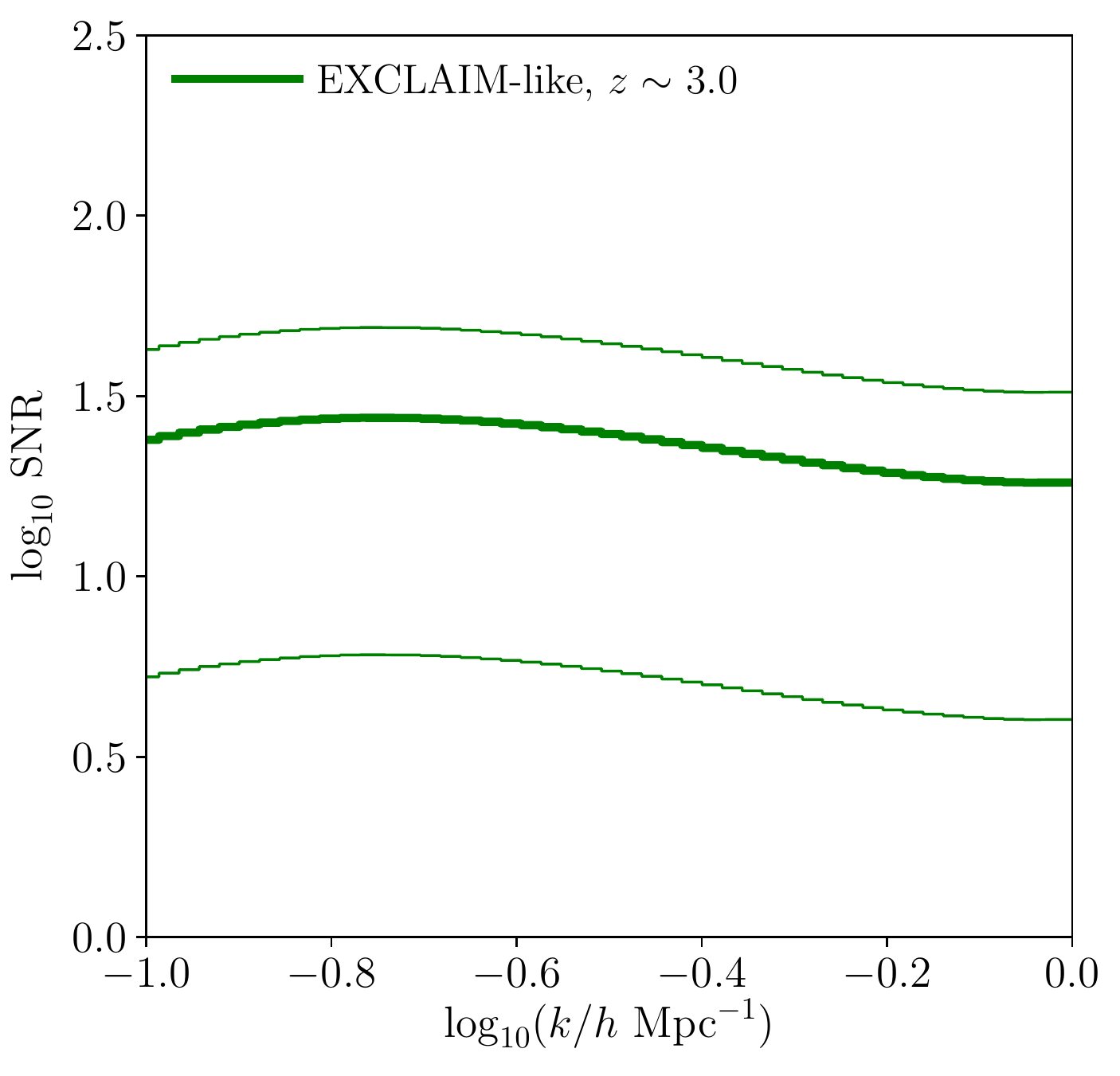} 
\caption{Signal-to-noise ratio at $z \sim 3$ from the empirical relation and 
the parameters of the EXCLAIM-like configuration. The associated uncertainties 
in the mean value of the signal are indicated by the thinner steps.}
\label{fig:snr3}
\end{figure}

 \begin{table*}
\begin{center}
    \begin{tabular}{ | c | c | c | c | c | c | c | p{1.5cm} |}
    \hline
    Configuration &  $D_{\rm dish}$ (m.) & $\Delta \nu$ (MHz) &  $N_{\rm spec, 
eff} $ & $S_{\rm A}$ (sq. deg.) & {{NEFD}} (mJy s$^{1/2}$) / $\sigma_{\rm N}$ 
(Jy s$^{1/2}$ / sr) & $B_{\nu}$ (GHz) & $t_{\rm surv}$ (h.) \\ \hline
    PIXIE-like & 0.55$^{\dagger}$ & 15000 & 4 & 20000 & $5.83 \times 10^6 
{\dagger}$ & 120 & 1500 \\
    STARFIRE-like & 2.5 & 1700  & 100 & 1 & $2.6 \times 10^7 {\dagger}$ & 15 & 
450 \\
    EXCLAIM-like & 0.74 & 1000 & 6 & 400 & $6 \times 10^5 {\dagger}$ & 120 & 8 
\\
    CONCERTO-like & 12 & 1500 & 1500 &  1.4 &155 & 40 & 1200 \\ 
    TIME-like & 12 & 400 & 32 & 0.01 & 65 & 40 & 1000 \\
   CCAT-p-like  &  6 & 400 &  200 & 16 & $2.5 \times 10^6 {\dagger}$  & 40  & 
4000 \\ 
     ALMACAL-like $^{*}$ & 12 & 15 & 3000 & 0.0014 & 2.85$^{*}$ & 8 & 1000 \\ 
\hline
    \end{tabular}
\end{center}
\caption{Various experimental configurations considered in this work.  
$^{\dagger}$: A FWHM of 1.65 deg is used as a good approximation for the 
PIXIE-like configuration \citep{switzer2017} instead of the usual diffraction 
formula, due to the detectors being highly multi-moded. For the PIXIE-like, 
STARFIRE-like, EXCLAIM-like and CCAT-p-like configurations, the value of 
$\sigma_{\rm N}$ is directly quoted in units of Jy s$^{1/2}$ / sr; for all 
other experiments, the value of the NEFD is quoted in mJy $s^{1/2}$. For PIXIE, 
the $\sigma_{\rm N}$  is calculated from \eq{sigmandef} assuming the NEP value 
$0.7 \times 10^{-16}$ W/$\sqrt{\rm Hz}$ and the etendu of 4 cm$^2$ sr from 
\citet{kogut2011}. For the STARFIRE-like experiment, the median value of 
$\sigma_{\rm N}$ from the experiment description in 
https://asd.gsfc.nasa.gov/conferences/FIR/posters/Aguirre\_STARFIRE.pdf is 
used. 
$^{*}$For the ALMA intensity mapping survey considered here, we use an 
ALMACAL-like configuration   
targeting [CII] in Band 6. The NEFD is calculated from the rms sensitivity in a 
field, which is taken to be roughly $S_{\rm rms} = 0.09$ mJy on combining 30 
pointings.}
 \label{table:ciisurveys}
\end{table*}

For each experiment, we calculate the noise power spectrum using the following 
equations \citep{silva2015, serra2016, dumitru2018}:
\begin{equation}
P_{\rm N} = V_{\rm pix} \frac{\sigma_{\rm N}^2}{t_{\rm pix}}
\label{noise}
\end{equation}

We now consider the terms in the RHS of the above \eq{noise} one by one.
The time per observing pixel, $t_{\rm pix}$ is given by:
\begin{equation}
t_{\rm pix} = t_{\rm obs} N_{\rm spec, eff} \frac{\Omega_{\rm beam}}{S_{A}}
\label{eqntpix}
\end{equation}

In the above expression, $t_{\rm obs}$ is the total observing time, $S_{A} $ is the survey area,  $\Omega_{\rm beam} = 2 \pi \sigma_{\rm beam}^2$, 
with $\sigma_{\rm beam} =  \theta_{\rm beam}/2.355$ in terms of the beam 
angular size $\theta_{\rm beam}$. The beam FWHM is calculated using the diffraction formula $\theta_{\rm beam} = 
1.22 \lambda_{\rm obs}/D_{\rm dish}$, where $\lambda_{\rm obs} = 158 \  \mu m 
(1 + z)$ for all the experiments except the PIXIE-like configuration. For 
PIXIE, we assume a FWHM of 1.65 deg due to the detectors being highly 
multi-moded \citep[][and E. Switzer, private communication]{switzer2017}. 
$N_{\rm spec, eff}$ is the effective number of detectors which integrate in 
parallel on voxels on a given frequency.

The pixel volume, $V_{\rm pix}$  is given by \citep{gong2012, dumitru2018}:
\begin{eqnarray}
V_{\rm pix} &=& 1.1 \times 10^3  {\rm (cMpc}/h)^3 \left (\frac{\lambda}{158  \ 
\mu m} \right) \left(\frac{1 +z}{8} \right)^{1/2}  \nonumber \\
&& \left(\frac{\theta_{\rm beam}}{10 '} \right)^2 \left(\frac{\Delta \nu}{400 
{\rm MHz}} \right)
\label{eqnvpix}
\end{eqnarray}
as a function of the rest wavelength $\lambda$ of the line transition, the 
redshift $z$ and the spectral resolution $\Delta \nu$. 

The variance per pixel, $\sigma_{\rm N}$ in \eq{noise} is given by:
\begin{equation}
\sigma_{\rm N} = \frac{\rm NEFD}{\Omega_{\rm beam}}
\label{sigmandef}
\end{equation}
where the NEFD (Noise Equivalent Flux Density), can be calculated on the sky 
for the instrument under consideration by using the expression:
\begin{equation}
{\rm NEFD} = \frac{\rm NEI}{\sqrt{N_{\rm detectors}}}
\end{equation}
where the NEI is the Noise Equivalent Intensity (in units of Jy s$^{1/2}$/sr) 
and represents the intensity required to achieve a signal-to-noise ratio of 
unity at the detector, and $N_{\rm detectors}$ is the number of detectors. 
\footnote{For some experiments, the value of the Noise Equivalent Power (NEP) 
is quoted instead of the NEI; the NEI is then calculated by using  NEI 
$\approx$ NEP/etendu where the etendu $A_{\rm det} \Omega_{\rm beam}$ is the 
product of the detector area $A_{\rm det}$ and the detector solid angle 
$\Omega_{\rm beam}$.}

For the ALMACAL-like configuration, the survey considered targets [CII] in Band 6 
using four 2 GHz windows. The NEFD = $S_{\rm rms} t_{\rm int}^{1/2}$ is 
calculated from the integration time $t_{\rm int} =1000$ hours and the the rms 
sensitivity $S_{\rm rms}$  in a field, which is taken to be roughly $S_{\rm 
rms} = 0.09$ mJy on combining 30 pointings for this survey.\footnote{Note that 
the product of $t_{\rm int}^{1/2}$ and $S_{\rm rms}$ remains constant given a 
particular configuration.} The $N_{\rm spec, eff}$ here corresponds to the 
number of spatial resolution elements, which is close to 3000 for the full 
survey area (R. Dutta and T. Mroczkowski, private communication).

Once the noise power is known, the variance of the [CII] observation is 
calculated as:
\begin{equation}
{\rm var}_{ \rm CII} = \frac{(P_{\rm CII} + P_{\rm N})^2}{N_{\rm modes}} 
\label{varcii}
\end{equation}
with the number of modes given by:
\begin{equation}
N_{\rm modes} = 2 \pi k^2 \Delta k \frac{V_{\rm surv}}{(2 \pi)^3}
\label{nmodes}
\end{equation}
with $\Delta k$ denoting the bin width in $k$-space (for all the experiments 
considered here, we use logarithmically equispaced $k$-bins with $\Delta \ln k 
= 0.05$), and the survey volume \citep{gong2012, dumitru2018} given by:
\begin{eqnarray}
V_{\rm surv} &=& 3.7 \times 10^7  {\rm (cMpc}/h)^3 \left (\frac{\lambda}{158  \ 
\mu m} \right) \left(\frac{1 +z}{8} \right)^{1/2}  \nonumber \\
&& \left(\frac{S_{A}}{16 {\rm deg}^2} \right) \left(\frac{B_{\nu}}{20 
{\rm GHz}} \right)
\label{vsurvey}
\end{eqnarray}
Finally, the SNR is calculated from \eq{varcii} as:
\begin{equation}
{\rm SNR} = \frac{P_{\rm CII}}{({\rm var}_{ \rm CII})^{1/2}}
\end{equation}
for each experiment under consideration.

\begin{figure}
\includegraphics[scale = 0.6, width = \columnwidth]{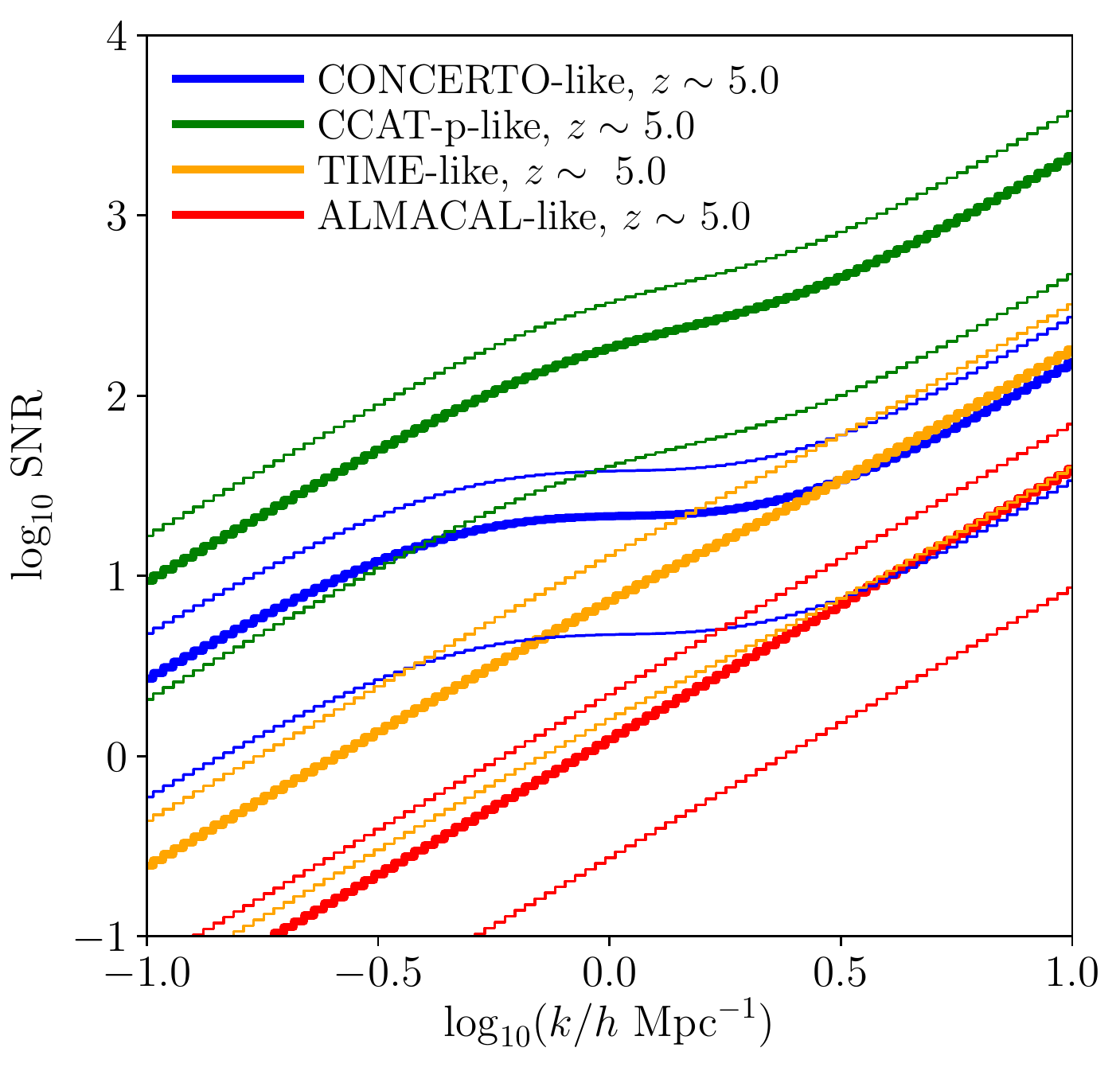} 
\includegraphics[scale = 0.6, width = \columnwidth]{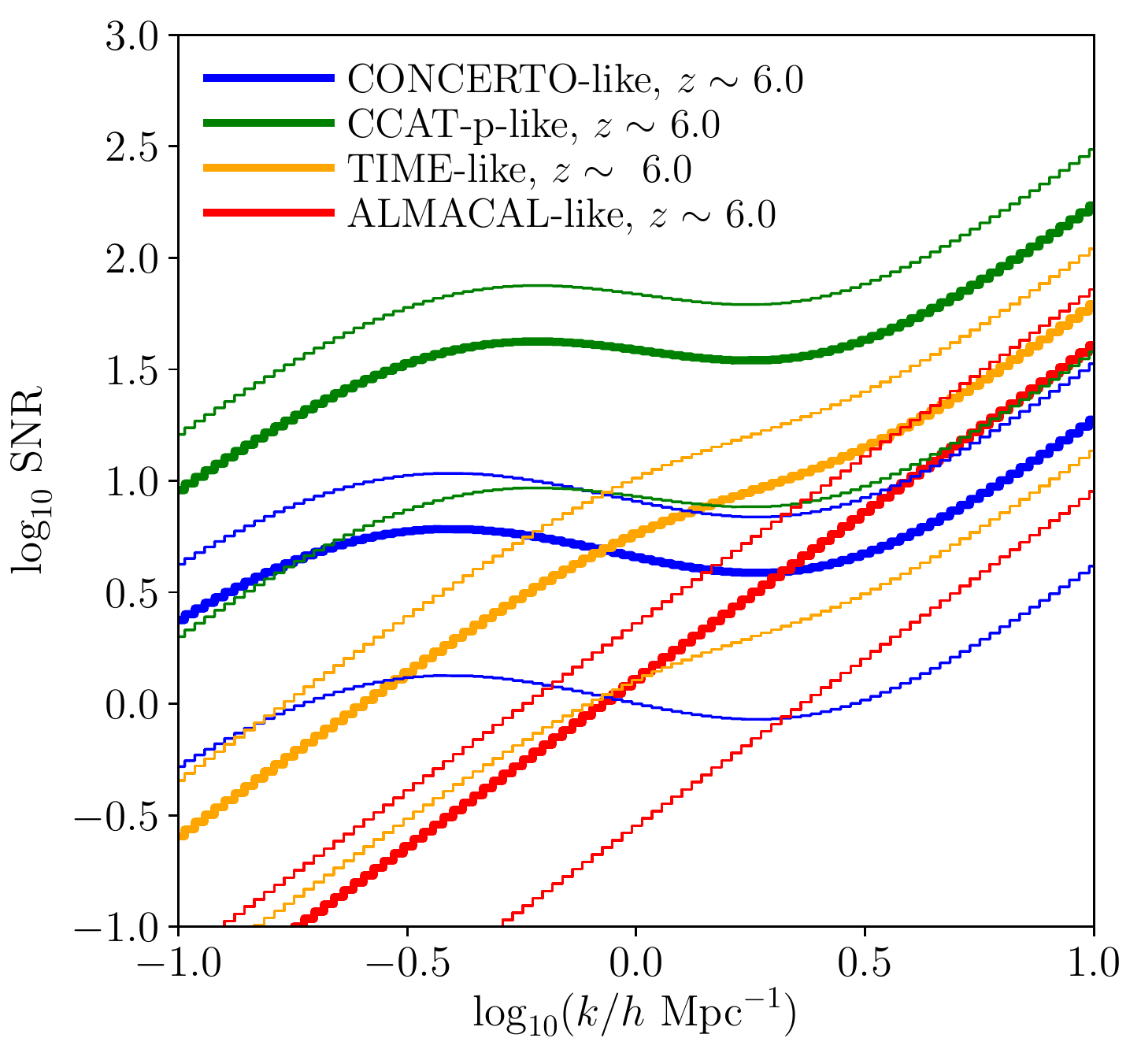} 
\caption{Signal-to-noise ratios at  $z \sim 5$ (top panel) and $z \sim 6$ 
(lower panel) from the empirical relation and the parameters of four future [CII] 
experiments: with the CCAT-p-like, CONCERTO-like, TIME-like and ALMACAL-like 
intensity mapping configurations. The associated uncertainties in the mean 
value of the signal are indicated by the thinner steps.}
\label{fig:snr56}
\end{figure}

The SNRs for surveys with the seven experiments described above and in Table 
\ref{table:ciisurveys} are plotted in the panels of Figs. \ref{fig:snr01}, 
\ref{fig:snr3} and \ref{fig:snr56} respectively. As the sensitivity depends on 
the assumed $k$-bin size ($\Delta k$) through the number of modes 
(\eq{nmodes}),  the SNRs for each experimental configuration are plotted as 
stair steps in log-$k$ space. The associated signal uncertainties are indicated 
by the thinner steps in each case. Note that the SNRs derived here may be 
somewhat optimistic since the derivation above does not take into account the 
beam and spectral width transfer functions. Specifically, SNR estimates beyond 
$k \sim 1$ Mpc$^{-1}$ may be subject to caveats coming from beam size in some 
configurations. {\footnote { \citet{chung2018a} adopt a more sophisticated line-scan survey strategy for TIME, finding a combined SNR across $k \lesssim 1$ to be of order unity at $z \sim 6$.  Noting that the water vapour levels at the telescope site considered in that work lead to an equivalent NEFD of about 100 (D. Chung, private communication), the SNR values here are consistent with \citet{chung2018a} to within an order of magnitude.}}

{To illustrate the scaling of the SNR with the parameters of the experimental configurations, we consider the examples of the CONCERTO-like and CCAT-p-like cases at $z \sim 6$.
From Eqs. (\ref{eqntpix}) and (\ref{eqnvpix}), we have:
\begin{equation}
\frac{V_{\rm pix}}{t_{\rm pix}} \propto \frac{\Delta \nu  S_{A}} {t_{\rm obs}  N_{\rm spec, eff}}
\end{equation}
   and  hence (using Table \ref{table:ciisurveys}):
\begin{equation}
  \left(\frac{V_{\rm pix}}{t_{\rm pix}} \right)_{\rm concerto} / \left(\frac{V_{\rm pix}}{t_{\rm pix}} \right)_{\rm ccatp} = 0.15
\end{equation}
Using \eq{sigmandef} to calculate $\sigma_{\rm N}$ for the CONCERTO-like case at $z \sim 6$, we find:
\begin{equation}
 \left(\sigma_{\rm N}^2 \right)_{\rm concerto}/\left(\sigma_{\rm N}^2\right)_{\rm ccatp}  \sim 18.72
 \end{equation}
which places the noise powers ($P_{\rm N}$'s) for the CONCERTO-like and CCAT-p-like configurations roughly a factor 2.81 apart.

Noting from \eq{vsurvey} that $V_{\rm surv} \propto B_{\nu} S_A$, we find that
\begin{equation}
 \left(V_{\rm surv} \right)_{\rm concerto}/\left(V_{\rm surv} \right)_{\rm ccatp}  = 0.0875
 \end{equation}
 which can be used in \eq{nmodes} to calculate the ratio of $N_{\rm modes}$ .
 
 When both the CONCERTO-like and CCAT-p-like surveys are in their noise dominated regimes, this leads to the required ratio of the SNRs:
 \begin{equation}
 {\rm SNR}_{\rm ccatp}/{\rm SNR}_{\rm concerto}  \sim 9.50
 \end{equation}
This is illustrated in the lower panel of Fig. \ref{fig:snr56}, which shows the mean SNRs of the CONCERTO-like and CCAT-p-like configurations spaced roughly an order of magnitude apart, at a typical scale of $k \sim 1$ where both are noise dominated. {\footnote { For the CONCERTO-like experiment, we find the SNR at $z \sim 6$ to be of 
order a few around $k \lesssim 1$, which is roughly consistent with the results of \citet{dumitru2018}, noting 
that our model predictions in Fig. \ref{fig:powspec56} around $z \sim 6$ are close to 
those in \citet[][Fig. 3 top panel]{dumitru2018}. Note that we 
assume in this case (D. Chung and G. Lagache, private communication) that the configuration uses 1500 pixels per array (one for 200-360 GHz and one for 125-300 GHz).}}

The `turnover' scale of the SNR at which a survey switches from being noise-dominated to sample variance (or cosmic variance) dominated occurs when $P_{\rm CII}  > P_{\rm N}$ in \eq{varcii}. This is relevant for configurations with smaller survey areas, e.g., in the ALMACAL-like configuration, the SNR  at $z \sim 5-6$ (where $P_{\rm CII} \gg P_{\rm N}$) is essentially a function of
the survey volume  throughout the $k$-range considered here:  ${\rm SNR} = N_{\rm modes}^{1/2}$. This also occurs in  the STARFIRE-like case  at $z \sim 1.5$ where the [CII] signal  is much larger than 
the noise term. }

\section{Conclusions}
\label{sec:conclusions}

We have developed a data-driven, halo model based framework towards 
interpreting future intensity mapping experiments involving the ionized carbon 
(CII) emission with rest-wavelength $158 \  \mu$m. Combining data from low 
redshift galaxy survey constraints \citep{hemmati2017} on the [CII] luminosity 
function, the empirical evolution of the star-formation rate density 
\citep{madau2014}  and the high-redshift constraints from the recent intensity 
mapping experiment at $z \sim 2.6$ \citep[][assuming the CII 
detection]{pullen2018}, we develop fitting forms for the local [CII] luminosity - 
halo mass relation and its predicted evolution with redshift. Although the 
current data at higher redshifts  mainly provide upper or lower limits to the 
[CII] luminosity function, the predicted evolution is fully consistent with these 
constraints. This formalism is used to predict the power spectra of [CII] 
intensity fluctuations both at intermediate redshifts as well as towards the 
end stages of reionization.  The best-fitting model and the parametrisation are 
summarized in Table \ref{table:final}.

As indicated in \citet{pullen2018}, a confirmation of this initial [CII] 
intensity mapping measurement would provide exciting insights for galaxy 
evolution and the metallicity of the ISM. From the results of the present work, 
we find that it can also be used to investigate how the [CII] - SFRD relation 
evolves with redshift during and after the peak of star formation in the 
universe. An assumption implicit in the present model is that the [CII]-SFRD 
connection continues to hold up to $z \sim 6$ representing the end stages of 
reionization.  This is supported by the recent findings of \citet{smit2018} who 
find no evidence for a deviation in the mean [CII] luminosity - SFR relation at 
$z \sim 6.8$ (compared to $z \sim 0$ and $z \sim 5.5$ galaxies) from ALMA 
observations of [OIII] - selected galaxies ; however, at higher redshifts ($z 
\sim 7$), other results \citep{pentericci2016} suggest a deficit in the [CII] 
luminosity compared to the lower redshift CII - SFR relation in Ly-$\alpha$ 
selected CII galaxies.  Clearly,  further constraints on this dependency would 
be possible with  the present model  in conjunction with future [CII] surveys in 
the late stages of the reionization phase of the universe. Larger, homogeneous 
CII galaxy detections would also help place independent constraints on the 
evolution of the slope of the [CII] - SFRD relation, thus  helping confirm the 
robustness of the intensity mapping measurements. Extrapolating this dependence 
to even higher redshifts would be possible with the data from blind or 
dedicated surveys with the JWST/ALMA.

The present work does not examine the effects of foregrounds, which would 
presumably be the limiting systematic in [CII] surveys at reionization epochs.  
The most dominant interloper lines arise from the CO (3-2) and (4-3) emission  
from $z \sim 0-2$, with an additional component coming from the cosmic infrared 
background (CIB).  Efficient removal techniques (cleaning/masking) have been 
developed  to mitigate line foregrounds \citep[e.g.,][]{breysse2015, cheng2016, 
visbal2010, lidz2016, sun2018}. At lower redshifts, the main contaminants are 
expected to be Galactic and extragalactic thermal dust emissions and their 
associated instrument response. However, at these frequencies, the results of 
simulations indicate excellent prospects for the robust recovery of the [CII] 
signal by using linear combination cleaning techniques 
\citep[e.g.,][]{switzer2017}.

It is of interest to explore cross-correlation possibilities with [CII] intensity 
mapping and large galaxy photometric/spectroscopic redshift surveys to be 
undertaken in the future, which would provide valuable information about the 
stellar and gas properties of dark matter haloes. Such cross-correlation 
studies in the context of CO  \citep[e.g.,][]{chung2018} may  promise good 
constraints on the astrophysical parameters of the CO luminosity - halo mass 
relation and would offer interesting complementary information in the case of 
[CII], especially in the context of planned surveys with  several future 
programs, e.g. the forthcoming ALPINE 
survey\footnote{http://cosmos.astro.caltech.edu/page/submm\#} which aims to 
measure [CII] properties in a sample of galaxies between $4 < z < 6$. 
Cross-correlations of galaxy survey data with [CII] intensity maps can also be 
used to shed light into various processes of galaxy formation and the metal 
enrichment of the ISM. Through the mid to end stages of reionization,  
synergies with HI and other surveys can offer exciting prospects into 
constraints on cosmology and astrophysics, including the baryon cycle and star 
formation rate, as well as the predicted sizes of ionization bubbles from the 
turnover of the cross-correlation coefficient \citep{visbal2010, 
gong2012,dumitru2018}.

\begin{table}
\centering
\caption{Summary of the best-fitting $L_{\rm CII} - M$ relation,  and the free 
parameters involved. The $L_{\rm CII}$ is in units of $L_{\odot}$ and all 
masses are in units of $M_{\odot}$.}
\label{table:final}
\begin{tabular}{|c|}
\hline
\\
$L_{\rm CII} (M, z) = F(z) [(M/M_1)^{\beta}  \exp (-N_1 /M)] $;\\ 
\\
$F(z) = ((1+z)^{2.7}/(1 + [(1+z)/2.9]^{5.6}))^{\alpha}$
 \\
\\
\hline 
\\
$M_{1} = (2.39 \pm 1.86) \times 10^{-5}$

; \  \\
\\ $N_{1} = (4.19 \pm 3.27) \times 10^{11} M_{\odot}$

           ; \ 
 \\
 \\
 $\alpha = 1.79 \pm 0.30$

 \\
\\$\beta = 0.49 \pm 0.38$

 \\

\hline
\end{tabular}
\end{table}

\section*{Acknowledgements}
I thank Marco Viero, Girish Kulkarni, Dongwoo Chung, Guochao Sun, Marta Silva, Patrick Breysse 
and Guilaine Lagache for helpful initial conversations and discussions related 
to [CII] intensity mapping. I thank Martin Zwaan, Gerg\"o Popping, Fabian Walter, 
Celine Peroux,  Rajeshwari Dutta and Tony Mroczkowski for useful discussions 
especially regarding possible ALMA intensity mapping surveys, and the ESO, 
Garching and the MPIA, Heidelberg for hospitality during my visits. I thank 
Simon Foreman, Anthony Pullen, Christos Karoumpis and Jos\'e Fonseca for 
detailed comments on the manuscript. I am grateful to Marta Silva, Sebastian 
Dumitru and Gerg\"o Popping for sharing data from their simulations. 
It is a pleasure to thank Dongwoo Chung for  a very careful reading of the 
manuscript, several useful comments and for sharing a draft version of his work 
in preparation. {I thank the referee for a helpful report that improved the presentation of the paper.} 
I also thank Guilaine Lagache, Eric Switzer, Celine Peroux and Tony Mroczkowski 
for helpful inputs regarding the various instrumental and survey designs used. 
My research was supported by the Tomalla Foundation.

\bibliographystyle{mnras}
\bibliography{mybib}

\label{lastpage}
\end{document}